\documentclass[pre,aps,floatfix,superscriptaddress,twocolumn]{revtex4}
\usepackage{graphicx}
\usepackage{dcolumn}
\usepackage{latexsym}
\usepackage{amsmath, amsthm, amssymb}
\usepackage{epsfig}
\usepackage{bm}
\usepackage[dvipsnames]{xcolor}

\begin{document}

\title{Flocking in one dimension: effect of update rules}

\author{R. K. Singh}
\email[]{rksingh@imsc.res.in}
\affiliation{The Institute of Mathematical Sciences, CIT Campus, 
4th Cross Street, Taramani, Chennai- 600113, India}

\author{Shradha Mishra}
\email[]{smishra.phy@itbhu.ac.in}
\affiliation{Department of Physics, Indian Institute of Technology(BHU), 
Varanasi- 221005, India}

\begin{abstract}
	In this study the effect of parallel and random-sequential updates on the dynamical properties 
	of flocks in one dimension is considered. It is found that the frequency of directional switching is 
	increased for random-sequential updates as compared to a parallel update. The nature of 
	disorder to order transition is also affected by the difference of updating mechanism: discontinuous 
	for parallel and continuous for random-sequential updates. 
\end{abstract}

\maketitle

\section{Introduction}
Collective motion is a ubiquitous phenomena observed in active systems 
driven out of equilibrium across widely separated length scales from 
single cells \cite{kemkemer}, to unicellular organisms \cite{bonner}, to 
bird flocks \cite{parrish}, and humans \cite{helbing}. Emergence of such a motion in a 
group of self-propelled units is termed as a flocking transition \cite{nature} and 
was reported by Vicsek and coworkers for a system of particles in two dimensions 
\cite{vicsek}. Although most of the natural systems of interest are generally two or 
three dimensional, emergence of collective motion in one dimension has attracted 
attention in recent times \cite{barabasi, solon}. Such one dimensional 
flocks exhibit the interesting property of direction switching \cite{evans, dossetti} 
and recent theoretical and experimental studies have proven the usefulness of the study of 
collective motion in one dimension, in particular relevance to the phenomena of directional 
switching \cite{yates, buhl}. 

Most such models studying flocking in one dimension generally employ discrete time evolutions 
of the system which are closer in nature to the sense of time as represented in digital simulations. 
However, it is known that on a digital time-scale a system of multiple particles can exhibit 
properties which are not a true representation of the original dynamical system as has been 
observed in equilibrium \cite{choi} and nonequilibrium systems \cite{blok}. Such differences 
in update rules have lead to the appearance of new universality classes in coupled map 
lattices \cite{marcq, rolf}. The observations motivate us to study the effects of different 
update rules on flocking dynamics in a collection of self propelled particles. In order to proceed 
with our goal, we introduce a system of active Ising spins moving in the unit interval $[0, 1]$ 
with constant speed $v_0$ which interact locally in a neighborhood of radius $r$. We find 
that the differences in update rules reflect in both the transient and steady-state properties 
of the flocks. The paper is organized as follows: 
in the next section we compare the two update rules followed by the conclusion. 

\section{Comparison of the two update rules}
We start with a collection of active spins moving in the unit interval $[0,1]$ with a fixed 
speed $v_0$ and periodic boundary conditions. The position $x^i$ of the $i$th spin evolves as:
\begin{align}
\label{dyn}
x^i_{n+1} = x^i_n + s^i_n v_0. 
\end{align}
with $s^i_n$ being the spin state at time $n$. A spin $i$ interacts with all other spins in 
the interval $[x^i_n-r, x^i_n+r]$ by flipping its orientation in 
accordance with the Metropolis algorithm \cite{barkema}. If $f$ is the net spin 
in the interaction radius $r$, then depending on the product $s^i_n f$ the spin flips certainly 
if the product is negative and with probability $\exp(-\beta s^i_n f)$ when the product is positive, where the inverse 
temperature $\beta$ measures global randomness. To study long-range order in the system, we define: 
\begin{align}
\label{order}
m = \frac{1}{N}\sum_i s^i\, 
\end{align}
as the average orientation of the system. The magnitude $|m|$ serves as an appropriate order-parameter 
and takes values in the interval $[0, 1]$ with 0 representing completely disordered state 
and 1 the state of complete long-range order. The above system of active spins can evolve either by a parallel 
update or by a random-sequential update which we now define. 

In a parallel update rule, $\forall ~i = 1, ..., N$, $s^i$ is modified to $\tilde{s}^i$ based on the local interactions 
of each $s^i$. Position vector of the system then evolves according to: ${\bf x}_{n+1} = {\bf x}_n + v_0 \tilde{{\bf s}}_n$, 
followed by $\tilde{{\bf s}}_n \rightarrow {\bf s}_n$. 
In a random-sequential update, on the other hand, an active spin $i$ is chosen at random from the collection 
of $N$ spins and its spin $s^i$ is modified according to the Metropolis algorithm. The difference lies in the 
step that the updated value of spin $s^i$ is 
used immediately to modify the position of the $i^{th}$ particle according to 
(\ref{dyn}). This process is repeated $N$ times so that each spin gets an equal chance 
of update and this process of $N$ random flips constitutes one unit of time equivalent 
to a parallel update of $N$ spins. It is evident from the definition of the two update rules 
that a parallel update is synchronous whereas a random-sequential update is intrinsically 
asynchronous, as there is a randomness inherent in the very nature of the update rule. 
Such randomness has implications in the dynamics of the spin system and reflects 
in both the transient and steady-state properties. 

At $n = 0$, positions of the spins are chosen uniformly from the unit interval. Initial spin states are also 
chosen $\pm 1$  at random in all the following observations unless explicitly stated. 
\begin{figure}
\includegraphics[width=0.5\textwidth]{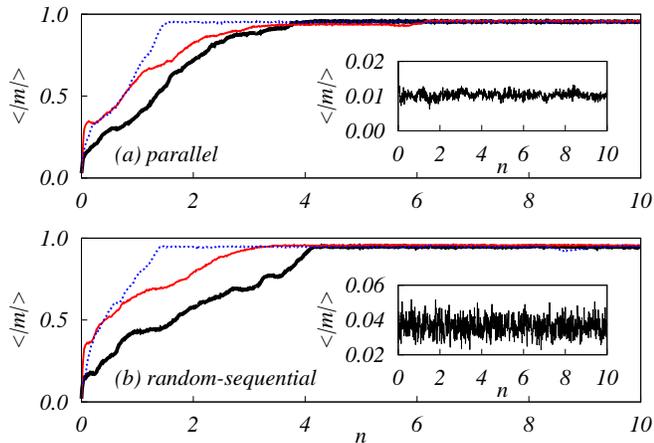}
\caption{Variation of the order-parameter $\langle |m| \rangle$ with time $n$(in multiples of 100) for a system 
	of $N = 1000$ spins for parallel(a) and random-sequential(b) updates. Parameter values for the 
	observations are: $(v_0, r) = $ (0.001, 0.01)(black), (0.001, 0.05)(red) and (0.003, 0.01)(blue) 
	respectively for inverse temperature $\beta = 4$. The inset in the two panels show the time-dependence 
	of $\langle |m| \rangle$ for $\beta = 1$ and $(v_0, r)$ = (0.001, 0.01) for the two update rules. 
The data are averaged over 30 ensembles. }
\label{fig1}
\end{figure}
Fig.~\ref{fig1} shows the variation of the order-parameter $\langle|m|\rangle$ with time $n$ for a 
system of $N = 1000$ spins moving in the unit interval $[0,1]$ for parallel(a) and random-sequential(b) 
updates respectively. For large inverse temperature, e.g.- $\beta = 4$, when the system exhibits long-range 
order, the $\langle |m| \rangle$ vs $n$ curves are similar for the two update rules: increased local 
interactions lead to a reduced time to achieve long-range order starting from complete disorder. 
However, for $\beta = 1$ we observe large fluctuations in the order-parameter for random-sequential updates 
as compared to that for parallel updates(inset in (a-b)). The fluctuations are intrinsic to the random-sequential 
update rule and later we will show that such fluctuations are prominent not only for low values of 
$\beta$ but also for higher values. 

\begin{figure}[h]
\includegraphics[width=0.5\textwidth]{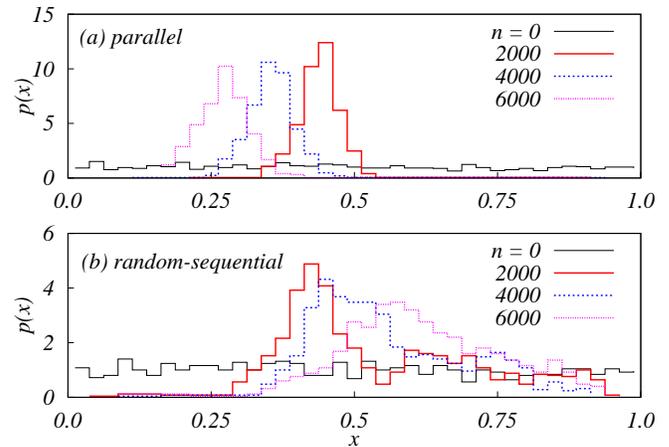}
\caption{Distribution of the positions $x$ of the spins for different snapshots of time $n$ 
	for parallel(a) and random-sequential(b) updates respectively. The distributions are 
	calculated over a single trajectory of the system starting with 
	a uniform distribution of spins over the unit interval at $n = 0$. The systems consist of 
	$N = 1000$ spins moving with speed $v_0 = 0.001$ at inverse temperature $\beta = 4$. 
	Interaction radius of the spins $r = 0.01$. 
}
\label{fig2}
\end{figure}
Fig.~\ref{fig2} shows snapshots of the probability distribution of positions $x^i$ of the spins 
for the two update rules for $\beta = 4$. Starting with a uniform distribution of the positions 
of spins at $n = 0$(black), the spins tend to move in close neighborhoods when the system exhibits 
long-range order  implying that the emergence of long-range order for high $\beta$ values is a 
mean-field effect. The cause for this effect is the propagation of the local interaction amongst the spins 
across the interval due to the finite speed of movement of the spins, i.e., $v_0 > 0$. Such a propagation 
of the local interaction leads to the emergence of a long-range order when global fluctuations are 
less(high $\beta$ values). This is because for high $\beta$, the probability of flipping  against the majority 
$\exp(-\beta s^i_n f)$, is less at any instant $n$ and hence the alignment of all the spins along the interval 
is achieved. On the other hand, for low values of $\beta$, the enhanced magnitude of global fluctuations increases 
the chance of any given spin $s^i$ to flip against the majority. As a result, even when the spins are moving 
with a constant speed $v_0$, a long-range order is not established because of the increased strength of 
global fluctuations which tends to disrupt the established local order at every instant. 
Now, random-sequential updates have an additional randomness due to the asynchronous updating mechanism which 
is reflected in the higher spread in comparison to the parallel counterpart. 
The clustered movement of the spins along the interval also implies towards the 
stability of the flocking state. For example, if a fraction of spins is flipped from their present state to 
reduce the value of the order-parameter $\langle |m| \rangle$, or in the extreme case, if the spins are completely 
randomized such that $\langle |m| \rangle \approx 0$, the flocking state of the system is restored to 
the previous value of $\langle |m| \rangle$. In addition, the time taken for the restoration of the flocking state 
after perturbation is less as compared to the time taken from $n = 0$. The reason for this reduction in 
time to restoring the flocking state is the proximity of the spins at the time of destabilization. 

\begin{figure}[h]
	\includegraphics[width=0.5\textwidth]{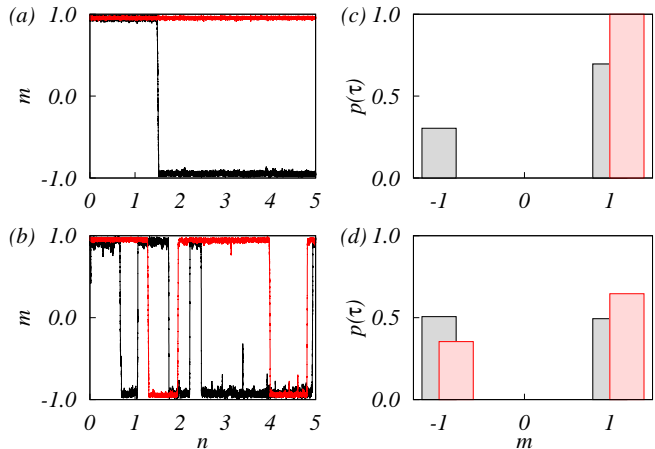}
	\caption{Variation of the order-parameter with sign $m$ with time $n$(in multiples of 
		$10^4$). Parts (a) and (b) represent a single trajectory of $m$ vs $n$ for parallel and 
		random-sequential updates respectively for $N = 1000$(red) and $N = 500$(black) spins 
		with $v_0 = 0.001$, $r = 0.01$ and $\beta = 4$. Starting with $m = 1$, 
		(c) and (d) represent the fraction of the 
		residence-time($\tau$) spent by the sample trajectories in the $m = 1$ and $m = -1$ states. 
		The histograms for $N = 1000$(red) and $N = 500$(black) are calculated by using data over 30 ensembles 
		and shifted relative to each other for clarity. 
	}
	\label{fig3}
\end{figure}
Next we report the directional switching behavior of the flocking state for parallel and random-sequential updates in 
Fig.~\ref{fig3}. Starting with the initial state in which all the spins are in $s^i = 1$ state, we find that 
the average spin $m$ fluctuates between $m = 1$ and $m = -1$ states for the two update rules for $\beta = 4$. 
The frequency of 
such flipping is, however, dependent on the size $N$ of the system as well as on the nature of update. For example, 
using the data for the evolution of the system over 50000 iterations and 30 ensembles, we find that for a parallel update 
rule the $N = 500$ size system fluctuates 60 times between the two states but no flipping is observed for $N = 1000$. 
On the contrary, for a random-sequential update the $N = 500$ size system exhibits 290 flips which is reduced to 
146 for $N = 1000$. In parts (c-d) of the Fig.~\ref{fig3} we report the residence-time $\tau$ statistics of the two states. 
The probability $p(\tau)$ of the residence-time of a given $m$ vs $n$ trajectory 
in $m = 1$ and $m = -1$ states shows that in the long-time 
limit, the trajectories tend to spend longer times in their initial state $m = 1$ for parallel updates whereas for a 
random-sequential updates the trajectories exhibit a true bistable behavior. These observations are a derivative of the 
local interactions and the intrinsic differences of the two update rules. 
The reduction in the alternating frequency of steady-state values of $m$ with increasing $N$ 
is consequent of the increased magnitude of local interactions against the same intensity of global noise $\beta$. 
In addition, the differences of the flipping frequency for the two update rules is attributed to the intrinsic 
fluctuations in the random-sequential updates. Our observation of the alternating steady-states is similar to 
a previous study for a lattice based model in one dimension \cite{evans}. 

\begin{figure}[h]
\includegraphics[width=0.5\textwidth]{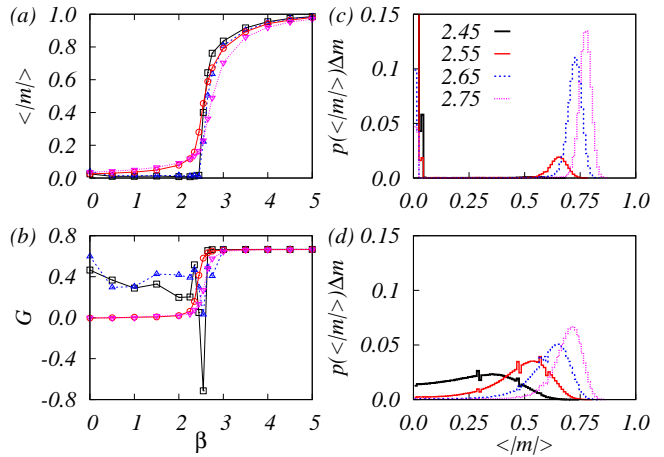}
\caption{Comparison of the steady-state properties for parallel vs random-sequential updates. 
(a-b) show the variation of the order-parameter $\langle|m|\rangle$ and the Binder cumulant $G$ in the steady state for 
parallel and random-sequential updates respectively. The steady-state properties are calculated using 
20000 iterations with 5000 iterations for the burn-in period and averaged over 10 ensembles. In parts (a) and (b), 
the symbols represent: parallel updates for $N = 500$(blue triangles) and $N = 1000$(black squares); and random-
sequential updates for $N = 500$(magenta inverted triangles) and $N = 1000$(red circles). 
(c) and (d) show the distribution of the order-parameter $|m|$ in the neighborhood of transition 
$\beta$ for four different values for the two update rules.  
The bimodal nature of the order-parameter distribution(c) and unimodal character with 
mean shifting towards right with increasing $\beta$(d) are characteristic of discontinuous 
and continuous transitions respectively. The distributions are calculated using 100000 iterations 
over 10 ensembles. All the calculations are done with $v_0 = 0.001$ and $r = 0.01$ and the probabilities of the 
order-parameter $p(\langle |m| \rangle) \Delta m$, $\Delta m = 0.01$ being the bin-width, are for a system 
of $N = 1000$ spins. }
\label{fig4}
\end{figure}
The steady-state properties of the system are shown in Fig.~\ref{fig4} depicting the nature of the transitions. 
It is observed that the nature of the flocking transition is different for the two update rules: 
with a first-order transition for parallel update and second-order for random-sequential updates. 
These are reflected in the $\langle |m| \rangle$ vs $\beta$ curves and the variation of the Binder cumulant 
$G = 1-\langle |m|^4 \rangle/3 \langle |m|^2 \rangle^2$ against the inverse temperature. A jump in 
$\langle |m| \rangle$ and  the strong negative 
values taken by $G$ imply that the disorder-to-order transition is first-order for parallel updates whereas 
it is second-order for random-sequential updates. It is to be noted that for 
  random-sequential updates $G$ varies smoothly 
  from {\it zero} to $2/3$ as the system goes from disordered state (small $\beta$) to long-ranged-ordered 
  state (large $\beta$), a consequence of the Gaussian nature of fluctuations of $\langle |m| \rangle$ about the mean. 
  But for parallel update $G >0$ in the disordered state (small $\beta$) and goes to $2/3$ values for long-ranged 
  ordered state (large $\beta$) with a strong negative value close to critical $\beta$. Although the mean 
  magnetization for parallel update is {\it zero} in the disordered state, $+ve$ value of $G$ arise due to the 
  deviation of the distribution of $\langle |m| \rangle$ from Gaussian about mean $\langle |m| \rangle = 0$.  
We also calculate the distribution of
the order-parameter $|m|$ in the neighborhood of transition $\beta$
for four different values for the two update rules 
(parallel(c) and random-sequential(d) respectively). The 
distributions show their respective properties of bimodality and unimodality which are 
characteristic of the two natures of transition: discontinuous and continuous. 

\section{Conclusions}
We have studied flocking in one dimension using a collection of active Ising spins 
moving in the unit interval. We find that the dynamical properties of the system: both 
transient and steady-state are intrinsically related to the nature of the update rules 
used to simulate the system. The magnitude of fluctuations in the disordered state of the spin system is 
more for random-sequential updates as against parallel updates. In the state of long-range order, 
the flocks alternate between the allowed orientations for the two update 
rules, the frequency of which is dependent on the strength of local interactions as well 
the type of update. For a fixed strength of local interactions, systems with parallel updates are less 
alternating in comparison to its random-sequential counterpart. The differences in the update rules 
also reflect in the transition from disorder to long-range order, with discontinuous for 
parallel updates whereas continuous for random-sequential updates. 
The differences arise due to intrinsic randomness in the random-sequential update 
which makes such an evolution asynchronous as opposed to parallel update which is 
inherently synchronous. The present study has implications towards the current understanding 
of collective motion in one dimension, in particular their modeling on digital computers.

\end{document}